\documentclass[aps,twocolumn,showpacs,amsmath,amssymb,nofootinbib]{revtex4} 
 
\usepackage{graphicx}
\usepackage{dcolumn}
\usepackage{bm}

\newcommand{\Npart}{N_{\text{part}}} 
\newcommand{\Tdec}{T_{\text{dec}}} 
\newcommand{\MeV}{\rm{~MeV}} 
\newcommand{\GeV}{\rm{~GeV}} 
\newcommand{\fm}{\rm{~fm}} 
\newcommand{\mb}{\rm{~mb}}

\begin{document}

\title{Elliptic flow of thermal photons in heavy-ion collisions at Relativistic Heavy Ion Collider and Large Hadron Collider} 
 
\author{H.~Holopainen} 
\email{hannu.l.holopainen@jyu.fi}
\author{S.~S.~R\"as\"anen}
\email{sami.s.rasanen@jyu.fi}
\author{K.~J.~Eskola}
\email{kari.eskola@phys.jyu.fi}
\affiliation{Department of Physics, P.O.Box 35, FI-40014 University of Jyv\"askyl\"a, Finland}
\affiliation{Helsinki Institute of Physics, P.O.Box 64, FI-00014 University of Helsinki, Finland} 
 
\begin{abstract} 
We calculate  the thermal photon transverse momentum spectra and elliptic flow in $\sqrt{s_{NN}} = 200 \GeV$ Au+Au collisions at RHIC and in $\sqrt{s_{NN}} = 2.76$~TeV Pb+Pb collisions at the LHC, using an ideal-hydrodynamical framework which is constrained by the measured hadron spectra at RHIC and LHC. The sensitivity of the results to the QCD-matter equation of state and to the photon emission rates is studied, and the photon $v_2$ is discussed in the light of the photonic $p_T$ spectrum measured by the PHENIX Collaboration. In particular, we make a prediction for the thermal photon $p_T$ spectra and elliptic flow for the current LHC Pb+Pb collisions. 
\end{abstract} 
 
 
\maketitle 

\section{Introduction}
Experimental data at the Relativistic Heavy Ion Collider (RHIC) and now also at the Large Hadron Collider (LHC) have shown compelling evidence of strongly interacting QCD-medium production in ultra-relativistic heavy-ion collisions. 
The measured transverse energies, transverse momentum ($p_T$) spectra and, in particular, the significant azimuthal anisotropy (elliptic flow) of final-state hadrons suggest together that partonic QCD matter, quark-gluon plasma (QGP), is formed in these collisions. 

In the collision of two nuclei, the azimuthally anisotropic overlap region sets preferred directions for the transverse flow. In hydrodynamical models, pressure gradients turn the spatial anisotropy  of the produced hot matter into a flow anisotropy, which is transmitted into the momentum distributions of measurable final-state hadrons at the decoupling of the system. However, the hadronic measurement reflects the flow (and temperature) conditions only in the freeze-out region where the hadronic interactions cease.

In comparison with partons, photons interact only very weakly with the QCD matter and thus a photon emitted
from the medium most likely escapes from the system without interacting. This is seen also in the measurements where photons do not show a similar suppression as hadrons when we move from proton+proton (p+p) to nucleus+nucleus ($A+A$) collisions \cite{Adler:2006hu}. Since photons can escape from the medium without interacting, they carry
information about the system at the time of their production.

Ultra-relativistic heavy-ion collisions are particularly interesting in regard with direct photon production, since relative to p+p collisions there are different types of nuclear effects at work as well as a number of important further sources for photons. In p+p (and also in p+$A$) collisions the direct photons are prompt photons originating from the primary hard interactions of partons, and fragmentation photons emitted by the primarily produced high-$p_T$ partons \cite{Aurenche:2006vj}. In heavy-ion collisions (and also in p+$A$), both the prompt and fragmentation photons at high-$p_T$ are subjected to nuclear effects in the parton distribution functions of the colliding nuclei (see e.g. \cite{Arleo:2011gc,Eskola:2009uj} for the quantification of these effects and their uncertainties). The fragmentation photon component is, however, expected to be suppressed due to the quenching of partonic jets in QCD matter in $A+A$ collisions. In addition to this, in $A+A$ collisions the jet(parton)-matter interactions, i.e. the jet-photon conversion \cite{Fries:2002kt,Fries:2005zh} and collision-induced photon emission from high-energy partons can produce photons which are important in the mid-$p_T$ region \cite{Turbide:2007mi,Liu:2008eh,Qin:2009bk}. Finally, the hot medium itself emits thermal photons, which are expected to be important in the few-GeV region and below, as discussed in the hydrodynamical studies of 
Refs.~\cite{Huovinen:1998tq,Huovinen:2001wx,Rasanen:2002qe,d'Enterria:2005vz,Turbide:2007mi,Liu:2008eh}.

In heavy-ion collisions, it is very difficult to distinguish between the different direct photon sources. In addition, there is a huge decay-photon background to deal with. The elliptic flow of direct photons could, however, shed more light on the interplay of the various photon production sources which differ from each other as follows: At high-$p_T$ (above $\sim 5$~GeV at RHIC), where prompt photons dominate \cite{Turbide:2007mi}, and where the fragmentation photons are more suppressed in the out-of-plane direction (perpendicular to the impact parameter), the photonic $v_2$ should be positive but very small \cite{Qin:2009bk}. The jet-medium interactions in turn increase the photon production most strongly in the in-plane direction (parallel to the impact parameter), thus causing a negative $v_2$ contribution at mid-$p_T$ \cite{Qin:2009bk}. The thermal photon production is affected by the hydrodynamical transverse flow itself, so that photons in the in-plane direction get a stronger boost. As shown earlier in Refs.~\cite{Chatterjee:2005de,Chatterjee:2008tp,Liu:2009kta}, this results in a positive elliptic flow for the thermal photons. Since the net contribution from other sources to photon $v_2$ 
is expected to be very small or even negative \cite{Qin:2009bk}, a large (hadron-like) photon $v_2$ measured in the few-$p_T$ region and below, should thus serve as a signature of thermal photon dominance. Since QCD matter is emitting photons throughout its entire evolution, measuring thermal photon $p_T$ spectra and $v_2$ would thus give important further constraints for the dynamics and properties of QCD matter.

In this work we focus on computing the thermal photon $p_T$ spectra and elliptic flow in $\sqrt{s_{NN}} = 200 \GeV$ Au+Au collisions at RHIC and in $\sqrt{s_{NN}} = 2.76$~TeV Pb+Pb collisions at the LHC, using an ideal-hydrodynamical framework which is constrained by the measured hadron spectra at RHIC and LHC. We study the sensitivity of the results to the QCD-matter equation of state (EoS) and to the photon emission rates.  We discuss the photon $v_2$ in the light of the photonic $p_T$ spectrum measured by the PHENIX Collaboration \cite{Adler:2006yt,Adler:2005ig,:2008fqa}. In particular, we  make a prediction for the thermal photon $p_T$ spectra and elliptic flow for the current LHC heavy-ion collisions. Previous predictions for the thermal photon $p_T$ spectra in Pb+Pb collisions at the planned maximum cms-energy 5.5 TeV of the LHC can be found in \cite{Arleo:2004gn,Dusling:2009ej}.

\section{Theoretical framework}

\subsection{Centrality classes}
Centrality classes for $A+A$ collisions studied here are calculated using the optical Glauber model. For nuclear
densities we use spherically symmetric Woods-Saxon profiles with the the thickness parameter $d = 0.54 \fm$ and radii $R_{\rm Au} = 6.37 \fm$ and $R_{\rm Pb} = 6.49 \fm$. The total cross section for $A+A$ collisions is calculated from
\begin{equation}
  \sigma_{\text{tot}}^{AA} = \int d^2\bm{b} \frac{d\sigma_{\text{tot}}}{d^2\bm{b}}
  = \int d^2 \bm{b} \Big( 1 - e^{-T_{AA}(\bm{b}) \sigma_{NN}^{\text{in}}} \Big),
\end{equation}
where $T_{AA}$ is the standard nuclear overlap function and
$\sigma_{NN}^{\text{in}}$ is the inelastic nucleon-nucleon cross section. We
take $\sigma_{NN}^{\text{in}} = 42 (64) \mb$ for $\sqrt{s_{NN}} = 200\, (2760)\GeV$.

The centrality classes are defined with impact parameter ranges $[b_i,b_{i+1}]$
so that for the centrality class of $c_i$ we have
\begin{equation}
  c_i = \frac{1}{\sigma_{\text{tot}}^{AA}} \int_{b_i}^{b_{i+1}} d^2 \bm{b}
        \Big( 1 - e^{-T_{AA}(\bm{b}) \sigma_{NN}^{\text{in}}} \Big).
\end{equation}
The average impact parameter for each centrality class is calculated using the
distribution $d\sigma/d^2\bm{b}$ as a weight. The average number of participants is
calculated similarly.  The obtained centrality classes, impact parameter ranges, average
impact parameters and number of participants are listed in
Table~\ref{tab: centrality}.

\begin{table}[bh]
  \begin{tabular}{ccccc}
    \hline
    \hline
    & centrality \% & $b$ range [fm] & $\langle b \rangle$ [fm] & $\Npart$ \\
    \hline
    & 0-5   & 0.00-3.35 & 2.24 & 346 \\
    & 5-10  & 3.35-4.74 & 4.08 & 289 \\
    & 10-15 & 4.74-5.81 & 5.30 & 242 \\
    & 15-20 & 5.81-6.71 & 6.27 & 202 \\
    & 20-30 & 6.71-8.21 & 7.49 & 153 \\
    RHIC & 30-40 & 8.21-9.49 & 8.87 & 102 \\
    & 40-50 & 9.49-10.6 & 10.1 & 64.4 \\
    & 50-60 & 10.6-11.6 & 11.1 & 37.5 \\
    & 0-20  & 0.00-6.71 & 4.47 & 267 \\
    & 20-40 & 6.71-9.49 & 8.18 & 128 \\
    \hline
    & 0-5  & 0.00-3.53 & 2.35 & 375 \\
    LHC & 0-20  & 0.00-7.05 & 4.70 & 294 \\
    & 20-40 & 7.05-9.98 & 8.60 & 141 \\
    \hline
    \hline
  \end{tabular}
  \caption{Various centrality classes for Au+Au collisions at $\sqrt{s_{NN}} =
           200 \GeV$ and for Pb+Pb collisions at $\sqrt{s_{NN}} = 2.76$~TeV, 
           obtained via the optical Glauber model.}
  \label{tab: centrality}
\end{table}

\subsection{Initial states for hydrodynamical evolution}

For RHIC we use the EKRT saturation model \cite{Eskola:1999fc} to
fix the initial entropy in most central collisions. As shown in
\cite{Eskola:2005ue,Niemi:2008ta} we can get a good description of the pion
spectra and the elliptic flow with this pQCD + saturation + hydrodynamics
approach. For Au+Au collisions at $\sqrt{s_{NN}} = 200 \GeV$ the model gives an
initial time $\tau_0 = 0.17 \fm$. For the $\sqrt{s_{NN}} = 2.76$~TeV Pb+Pb collisions at the LHC, we fix the initial entropy so that we reproduce the measured multiplicity \cite{Aamodt:2010pb}.  The initial time $\tau_0 = 0.12 \fm$ is based on the EKRT-motivated fit done in Ref.~\cite{Renk:2011gj}.

\begin{figure}[t]
  \includegraphics[height=9.0cm]{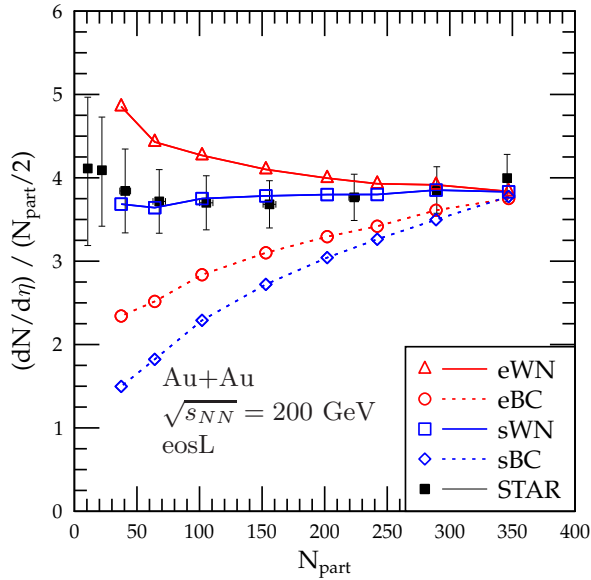}
  \caption{\protect\small (Color online) Number of charged hadrons at
           midrapidity in $\sqrt{s_{NN}} = 200 \GeV$ Au+Au collisions scaled
           by the number of the participant pairs calculated from the optical
           Glauber model. The data are from the STAR Collaboration \cite{:2008ez}.}
  \label{fig: charged mult}
\end{figure}

To fix the initial transverse density profile in $\sqrt{s_{NN}} = 200 \GeV$ Au+Au collisions, we do the following: 
In Fig.~\ref{fig: charged mult} we show, from Ref.~\cite{:2008ez}, the measured charged-particle multiplicity (divided by the number of participant pairs) as a function of the number of participants calculated from the optical Glauber model\footnote{Note that usually the number of participants quoted by the experiments here is from the MC Glauber model.}.
Choosing the initial transverse density according to the binary-collision-scaled energy or entropy density (eBC, sBC), or 
wounded-nucleon-scaled energy or entropy density (eWN, sWN) as introduced in Ref.~\cite{Kolb:2001qz}, we compute the charged-particle multiplicity in the centrality classes obtained above. The initial entropy at $b=0$ in these four cases is kept fixed. We see that the sWN profile fits the measured centrality dependence quite well. We will therefore choose the sWN profile at RHIC and, for simplicity, use the same profile also for the LHC Pb+Pb collisions. 

\subsection{Hydrodynamics and freeze-out}
To describe the spacetime evolution of the produced QCD matter, 
we solve the ideal-hydrodynamic equations
\begin{equation}
  \partial_\mu T^{\mu\nu} = 0,
\end{equation}
where $T^{\mu\nu} = (\epsilon + P) u^\mu u^\nu - P g^{\mu\nu}$ is the energy-momentum tensor, $u^\mu$ is the fluid four-velocity, $\epsilon$ is the energy density and $P$ is the pressure.
As we are interested in particle and photon production at midrapidity,
we may assume that net-baryon density is negligible. Since the particle
spectra are approximately flat at midrapidity, we can simplify our
hydrodynamical equations by assuming longitudinal boost-invariance. We use the
SHASTA algorithm \cite{Boris,Zalesak} to solve this (2+1)-dimensional
numerical problem.

To close the hydrodynamic equations we need an Equation of state (EoS), $P = P(\epsilon)$. In
this paper we study the sensitivity of thermal photon production to the EoS, by focusing on two different cases.
The first case, called here "eosQ", corresponds to the Bag model EoS with a first order phase transition \cite{Sollfrank:1996hd}. 
In eosQ, the high-temperature phase with the Bag constant is an ideal gas of three flavors of massless quarks and gluons, while the low-temperature phase is an ideal gas of all hadronic states with $m < 2$ GeV. These two phases are connected with a mixed phase, and the Bag constant is chosen so that the critical temperature is $T_c = 165 \MeV$.
The second EoS case, which we call "eosL", is adopted from Ref.~\cite{Laine:2006cp}. 
This EoS is quite similar to the recently constructed lattice EoS "s95-p" \cite{Huovinen:2009yb}, and as discussed in \cite{Huovinen:2009yb},  the hadron spectra and elliptic flow are in practice insensitive to the differences between eosL and s95-p. 

Since the lattice data suggests that the phase transition from the QGP to hadron gas (HG) is not of first order, one may consider the eosQ case somewhat unrealistic. However, the computation of thermal photon production in the phase-transition region requires well-defined QGP and hadron-gas fractions, which are available only in the eosQ case. 
With eosL, in the absence of such phase fractions, there are additional uncertainties in the thermal photon calculation  related to the QGP and HG emission rates.

Thermal transverse momentum ($p_T$) spectra of hadrons are obtained using the Cooper-Frye method
\cite{Cooper} where particle emission from a freeze-out hypersurface
$\sigma$ is calculated with
\begin{equation}
  \frac{dN^f}{d^2 p_T dy} = \int_\sigma f(x,p) p^\mu d\sigma_\mu,
\end{equation}
where $f(x,p)$ is the momentum distribution function of a specific hadron type. 
We assume the system to decouple at a single constant temperature $\Tdec$, 
which is fixed so that we get a good agreement with the measured
$p_T$ spectra of pions at RHIC. With eosQ, we have $\Tdec=140$~MeV, and 160 MeV with eosL .

After the thermal emission of particles from the freeze-out surface is calculated,
we take into account the strong and electromagnetic 2- and 3-body decays. This treatment is essential since
most of the stable particles in our case come from the decays of heavy resonances.

The $p_T$ spectra of hadrons can be written as a Fourier series,
\begin{equation}
 \frac{dN}{d^2 p_T dy} = \frac{1}{\pi} \frac{dN}{dp_T^2 dy} \Big( 1 + \sum_{n=1}^\infty 2 v_n \cos (n\phi) \Big),
\end{equation}
where $\phi$ is the hadron momentum's azimuthal angle with respect to
the reaction plane defined by the impact parameter. Elliptic flow, $v_2$, is the second
coefficient in this series and it can be computed from
\begin{equation}\label{eq: v_2}
  v_2(p_T) = \frac{ \int d\phi \cos(2\phi) \frac{dN(b)}{dp_T^2 d\phi dy} }{ \int d\phi \frac{dN(b)}{dp_T^2 d\phi dy} }.
\end{equation}
Correspondingly, the $p_T$-integrated $v_2$ becomes
\begin{equation}\label{eq: v_2_int}
  v_2 = \frac{ \int d\phi \cos(2\phi) \frac{dN(b)}{d\phi dy} }{ \int d\phi \frac{dN(b)}{d\phi dy} }.
\end{equation}

\begin{figure}[t]
  \includegraphics[height=9.0cm]{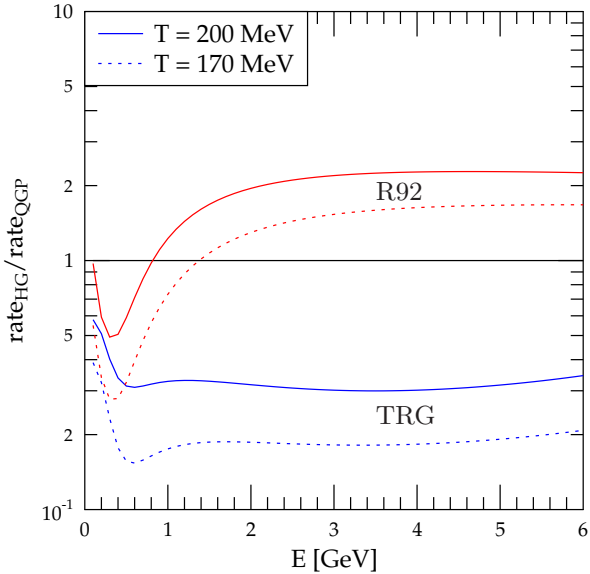}
  \caption{\protect\small (Color online) 
  The HG-to-QGP ratio of the photon emission rates as a function of the photon energy at
  two different temperatures. Two different HG-rates, R92 \cite{Kapusta:1991qp,Xiong:1992ui,Nadeau:1992cn} and TRG \cite{Turbide:2003si}, are compared.}
  \label{fig: rate comparison}
\end{figure}

\subsection{Thermal photons emission from the hydrodynamical medium}

The $p_T$ spectra  of thermal photons can be calculated from
\begin{equation}
  \frac{dN^{\gamma}}{d^2 p_T dy} = \int d^4x\, \Gamma (E^*(x),T(x)),
\end{equation}
where $\Gamma(E^*,T)$ is the Lorentz invariant thermal photon emission rate,
$d^4x$ is the volume element and $E^*(x)=p^\mu u_\mu(x)$ is the photon energy in the
fluid's local rest frame. For the QGP, we use the emission rate from Refs.~\cite{Arnold:2001ba,Arnold:2001ms} with $N_f=3$ and a running strong coupling constant \cite{Karsch:1987kz}
$\alpha_s = \beta /\ln(8 T/\Lambda)$, with $\beta=6\pi/(33- 2 N_f)$ and $\Lambda = 200$~MeV.
For the hadron gas, we use two different emission rates:
\textsl{(i)} Those calculated in \cite{Kapusta:1991qp,Xiong:1992ui} and parametrized
in \cite{Xiong:1992ui,Nadeau:1992cn}, which we call "R92". These rates  were used in the previously published LHC predictions \cite{Arleo:2004gn}.
\textsl{(ii)} The more recent ones from Ref.~\cite{Turbide:2003si} which account also for the finite size of hadrons through form factors. We call these rates "TRG".

With eosL, which smoothly goes from the QGP to the HG phase without specifying their volume fractions, one needs to choose how to switch from the QGP to HG photon emission rates. For simplicity, we choose to do this at a constant temperature $T_s$ but we vary $T_s$ between 170 and 200 MeV. We label these two cases as "eosL170" and "eosL200".

To illustrate the differences between the R92 and TRG emission rates which will be important for the photonic $v_2$ results presented in Sec. III ahead, we plot in Fig.~\ref{fig: rate comparison} the ratio of the photon emission rates
in the HG and QGP at two different fixed temperatures. We see that (since both HG rates have been divided by the same QGP rate) the difference between R92 and TRG is about a factor six at large energies. Furthermore, the TRG rates are always well below the QGP rates, while this is not the case for the R92 rates.

Elliptic flow for the thermal photons is calculated as in Eq.~\eqref{eq: v_2}.
Since the thermal photons cannot be distinguished from other direct photons, the elliptic flow from thermal photons alone cannot be measured. In what follows, we assume that the net contribution to the photonic $v_2$ from the other direct photon sources remains small, especially since the fragmentation photons with a positive
$v_2$ should partially cancel the negative $v_2$ of the photons arising from parton-medium interactions  \cite{Qin:2009bk}. 

If other components are emitted isotropically we can roughly estimate how much
they "wash away" the elliptic flow coming from thermal photons. The total
elliptic flow is then
\begin{equation}
\begin{split}
  v_2 &= \Big( \int d\phi \cos(2\phi) \frac{dN^{\text{th}}}{dp_T^2 d\phi dy} \Big) \Big( \int d\phi \frac{dN^{\text{all}}}{dp_T^2 d\phi dy} \Big)^{-1} \\
  &= v_2^{\text{th}} \Big( \frac{dN^{\text{th}}}{dp_T^2 dy} \Big) \Big( \frac{dN^{\text{all}}}{dp_T^2 dy} \Big)^{-1},
\end{split}
\label{eq:total_v2}
\end{equation}
where ${dN^{\text{all}}}/{dp_T^2 dy}$ corresponds to the measured  $p_T$ spectrum of direct photons and 
$v_2^{\text{th}}$ is the $v_2$ of thermal photons alone.

\begin{figure}[t]
  \includegraphics[height=8.0cm]{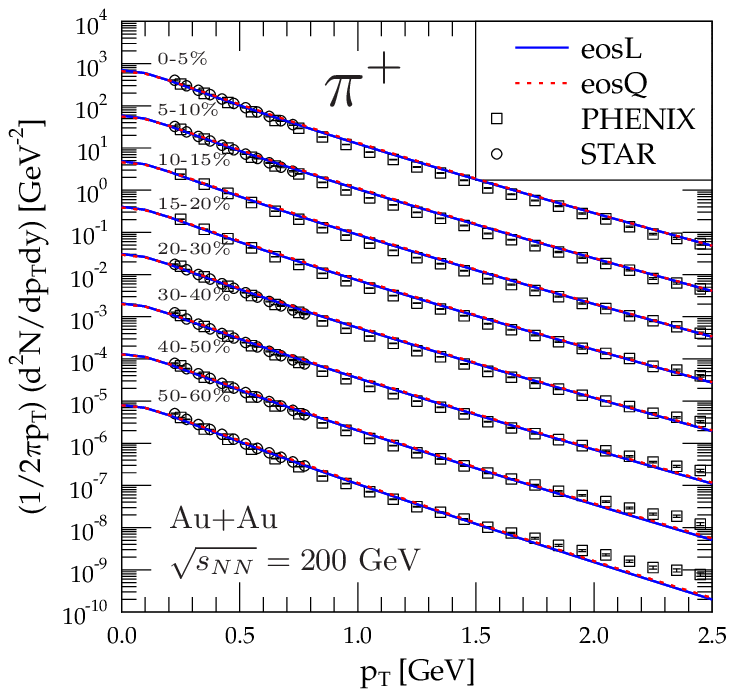}
  \caption{\protect\small (Color online) The $p_T$ spectra of positive pions
  for $\sqrt{s_{NN}}=200 \GeV$ Au + Au collisions at RHIC compared with the
  PHENIX data \cite{Adler:2003cb}. The centrality classes are indicated in
  the figure and the spectra are scaled by increasing powers of $10^{-1}$.}
  \label{fig: rhic pion spectra}
  \includegraphics[height=8.0cm]{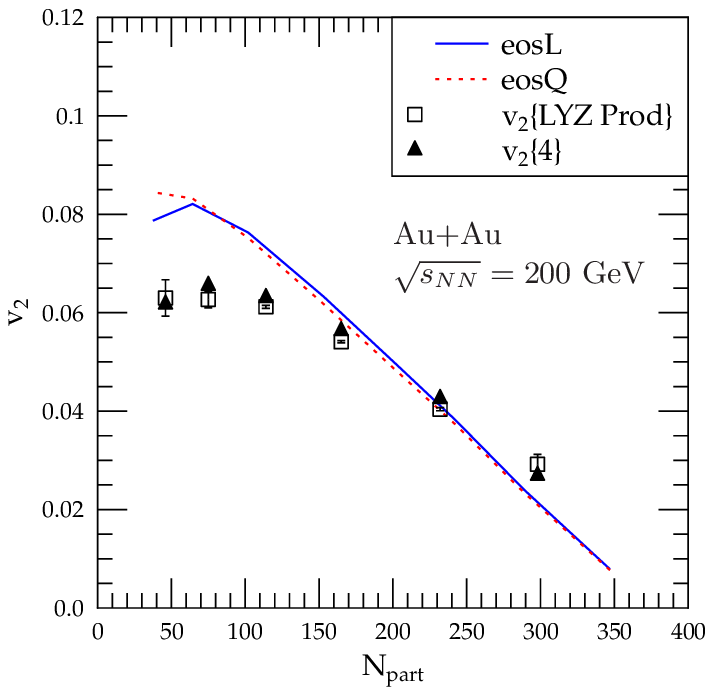}
  \caption{\protect\small (Color online) The integrated $v_2$ for charged
  hadrons in $\sqrt{s_{NN}}=200 \GeV$ Au + Au collisions at RHIC compared
  with the STAR data \cite{:2008ed}.}
  \label{fig: rhic v2 int}
\end{figure}

\section{Results for RHIC}

\subsection{Hadron spectra and elliptic flow}

First we show the hadronic observables to demonstrate that our hydrodynamical
description of the bulk QCD-medium is reasonable. Figure ~\ref{fig: rhic pion spectra}
shows the transverse momentum spectra of positively charged pions in different centrality classes in 
$\sqrt{s_{NN}} = 200$~GeV Au+Au collisions at RHIC. As we can see, we
have a good fit to the pion spectra below $p_T \approx 2 \GeV$ for a very wide range of
centralities. 
The integrated elliptic flow of charged hadrons from Eq.~\eqref{eq: v_2_int} is plotted in Fig.~\ref{fig: rhic v2 int} together with the data obtained by the STAR Collaboration \cite{:2008ed} using the 4-particle cumulant and LYZ methods which should best reflect the elliptic flow relative to the reaction plane defined by the impact parameter.
We have a fairly good description of the data also here, although the centrality dependence of the computed $v_2$ is not
fully reproduced. We expect, however, that fine-tuning the initial density profile, invoking the Monte Carlo Glauber model
and possibly also event-by-event hydrodynamics (see e.g. Ref.~\cite{Holopainen:2010gz}) as well as including viscous effects (see e.g. Ref.~\cite{Romatschke:2007mq}) will improve the agreement. These improvements are, however, beyond the scope of this exploratory paper. 
\begin{figure*}[t]
  \includegraphics[height=8.8cm]{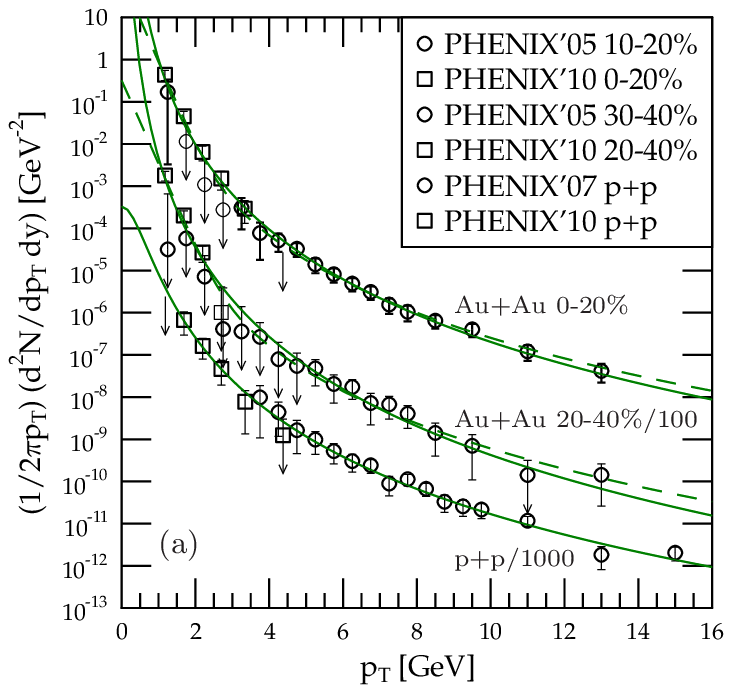}
  \includegraphics[height=8.8cm]{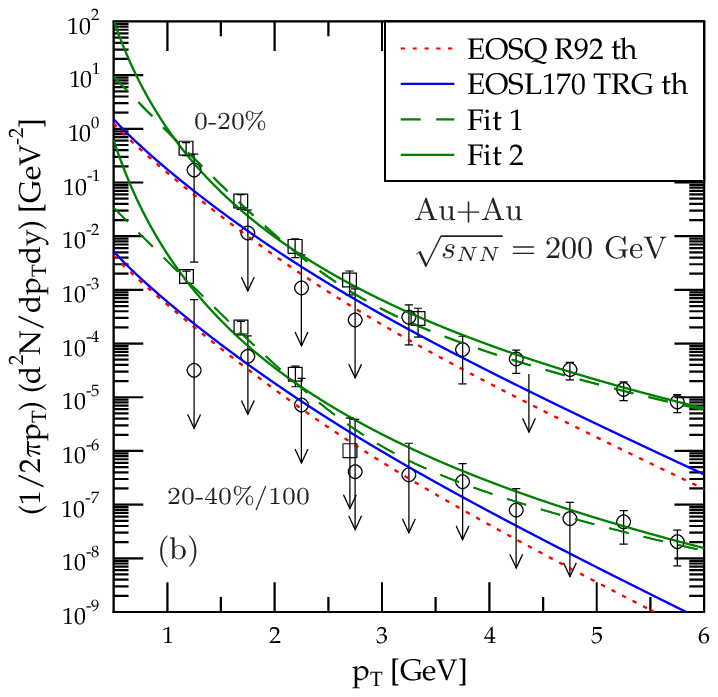}
  \caption{\protect\small (Color online) 
  The fits 1 and 2 to the measured photon spectra for two different centralities in $\sqrt{s_{NN}} = 200$~GeV Au+Au collisions  and in p+p collisions at $\sqrt{s} = 200$~GeV. The 20-40\% centrality-class spectra are divided by $100$ and  the p+p case by $1000$. The data are from the PHENIX Collaboration \cite{Adler:2006yt,Adler:2005ig,:2008fqa}. Our thermal photon results are shown in the panel (b). }
  \label{fig: photon spectra RHIC}
\end{figure*}

\subsection{Photon spectra}

In order to study how much elliptic flow is washed away by other sources
of direct photons at RHIC, we need to estimate the other components. We do this
by fitting the measured photon $p_T$ spectrum. To study the uncertainties due to the chosen fit functions, we
use two different forms. Our first choice ("fit 1") is an exponential
combined with a power law function \cite{:2008fqa}
\begin{equation}
  f(p_T) = A\exp(-p_T/T) + \frac{C}{(1 + p_T^2/b)^n},
\end{equation}
where $A,T,C,b$ and $n$ are the fit parameters. This fit function is physically
motivated by the QCD-like power-law behavior at high $p_T$ and the thermal-like exponential
at low $p_T$. Our alternative choice ("fit 2") is a mere power-law function
\begin{equation}
  f(p_T) = \frac{C}{(1 + p_T^2/b)^n}.
\end{equation}

In these fits, we use the photon data from PHENIX Collaboration. The older data
sets \cite{Adler:2006yt,Adler:2005ig} have large error bars at low $p_T$ but in the more
recent low-$p_T$ data \cite{:2008fqa} the error bars are much smaller. Unfortunately, the 
centrality classes in these measurements differ from each other. For our fits shown in 
Fig.~\ref{fig: photon spectra RHIC}, we have simply combined the 0-20\% (20-40\%) centrality data from Ref.~\cite{:2008fqa} with the 10-20\% (30-40\%) centrality data from Ref.~\cite{Adler:2005ig}. In our fits we have included all datapoints from the above sets. 

With the fit 1, we first find the parameters $b$ and $n$ by fitting the measured photon $p_T$ spectra in p+p-collisions using the PHENIX data \cite{Adler:2006yt,:2008fqa} shown in Fig.~\ref{fig: photon spectra RHIC}(a). Then for the Au+Au case, keeping the high-$p_T$ slope-parameters $b$ and $n$ fixed, we find $A$, $T$ and $C$ by fitting the PHENIX data \cite{Adler:2005ig,:2008fqa} for the two centrality classes shown in Fig.~\ref{fig: photon spectra RHIC}. For the fit 2, we use the same data sets. The best fit parameters obtained for the power law fits are listed in Table~\ref{tab: power}
and the parameters for the fit 1 can be found in Table~\ref{tab: exponential}. The fits 1 and 2 have equally small $\chi^2$ values at both centralities. 

\begin{table}[htb]
  \begin{tabular}{cccc}
    \multicolumn{4}{c}{Power law fit} \\
    \hline
    & C [GeV$^{-2}$]  & b [GeV$^{2}$]  & n \\
    \hline
    p+p & $3.29 \cdot 10^{-1}$  & $4.37 \cdot 10^{-1}$ & 3.09 \\
    Au+Au 0-20\% & $2.16 \cdot 10^{14}$ & $5.24 \cdot 10^{-5}$ & 3.35 \\
    Au+Au 20-40\% & $9.33 \cdot 10^{17}$ & $6.35 \cdot 10^{-6}$ & 3.52 \\
    \hline
  \end{tabular}
  \caption{\label{tab: power} The parameters obtained for the power law fits 2.}
\end{table}

\begin{table}[htb]
  \begin{tabular}{cccc}
    \multicolumn{4}{c}{Exponential + power law fit}  \\
    \hline
    & A [GeV$^{-2}$]  & T [GeV]  & C \\
    \hline
    Au+Au 0-20\% & 85.4  & 0.212 & 4.96 \\
    Au+Au 20-40\% & 30.7 & 0.218 & 1.18 \\
    \hline
  \end{tabular}
  \caption{\label{tab: exponential} The parameters for the exponential + power law fits 1.
  In this case, $n$ and $b$ are obtained from Table~\ref{tab: power}.}
\end{table}

In Fig.~\ref{fig: photon spectra RHIC}(b) we have replotted the low- and mid-$p_T$
region from Fig.~\ref{fig: photon spectra RHIC}(a), and shown our
thermal photon results obtained with eosL (eosQ) using the TRG (R92) rates in the
HG phase. For clarity, we have plotted the eosL results only for $T_s = 170$~MeV.
If we do the switch of the emission rate at $T_s = 200$~MeV, we
get 30\% (10\%) less photons at $p_T=1 (2)$~GeV, because the TRG emission
rate is smaller than the QGP emission rate, as shown in
Fig.~\ref{fig: rate comparison}.

Our thermal photon results shown in Fig.~\ref{fig: photon spectra RHIC}(b) differ by a factor
of two at high $p_T$. Some of this difference comes from the small difference in the initial temperature profiles which in our case are obtained from the fixed initial entropy density through the EoS.
However, a more dominant effect is the
different mapping of the energy density to the temperature in eosQ and eosL.
The difference in the actual temperature in the two cases is not large but the exponential temperature-dependence in the emission rates magnifies the effect considerably.

From the previous photon studies, see e.g. Refs.~\cite{Huovinen:2001wx,Rasanen:2002qe}, we know that photons from HG
are contributing mostly at small $p_T$. In Fig.~\ref{fig: photon spectra RHIC}
the difference between the eosQ+R92 and eosL+TRG results shrinks down at low $p_T$ 
since the R92 HG emission rate is larger than that in the TRG rates and since
with eosQ the HG volume becomes larger than with eosL.
When we use the TRG rates and eosL, only 3\% of the photons come
from the HG at $p_T = 1 \GeV$. With eosQ and the R92 rates about 50\% of the photons
originate from HG at the same $p_T$. 

We also note that in the low-$p_T$ region our thermal photon results very clearly undershoot the latest PHENIX data. We have checked that changing the freeze-out temperature to $\Tdec = 120 \MeV$ gives only a negligible improvement. This feature is typical to almost all hydrodynamical calculations as can been seen e.g. from
Fig.~43 in Ref.~\cite{Adare:2009qk}.

At $p_T \sim 3 \GeV$ the obtained thermal photon emission is almost enough to
match the fit 1 at both centralities if eosL is considered. This suggests
that we may have a window for thermal photon dominance at this $p_T$. However, if we
compare with the fit 2 there is always at least a factor of two difference.
Event-by-event fluctuations in the initial state, however,  have been shown to increase the thermal emission at
$p_T > 2 \GeV$ \cite{Chatterjee:2011dw}, and thus we should indeed have a better chance
to have a region where the direct photon $p_T$ spectrum, and consequently also the photon $v_2$, at RHIC is entirely dominated by the thermal emission.

\begin{figure*}[th]
  \includegraphics[height=8.8cm]{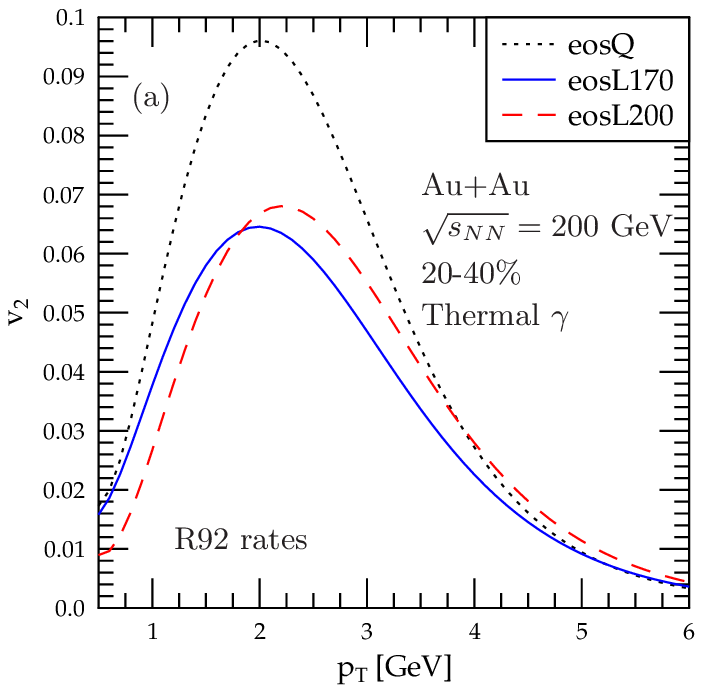}
  \includegraphics[height=8.8cm]{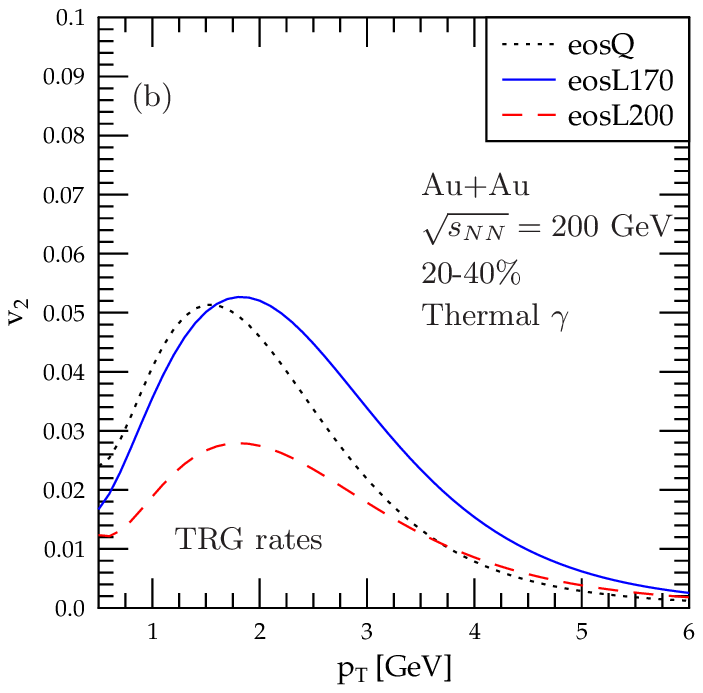}
  \caption{\protect\small (Color online) Elliptic flow of thermal
  photons in Au+Au collisions at $\sqrt{s_{NN}} = 200 \GeV$.}
  \label{fig: photon v2 thermal}
\end{figure*}

\subsection{Photon elliptic flow}

In Fig.~\ref{fig: photon v2 thermal}(a) we have plotted the elliptic flow of
the thermal photons using both eosQ and eosL with the R92 rates. 
Panel b shows the same calculations but with the TRG rates. Unlike for hadrons, the thermal photon elliptic flow
starts to decrease quickly above $p_T \sim 2 \GeV$. 
The reason for this is that practically all high $p_T$
photons are emitted nearly isotropically in the beginning of the evolution (see e.g. Fig.~3 in
Ref.~\cite{Chatterjee:2011dw}), when the hydrodynamical flow effects are very small.
Since the photon emission is dominated by the early times the
thermal photon elliptic flow is clearly smaller than the hadronic $v_2(p_T)$,
which probes the flow anisotropy only on the freeze-out surface.

From Fig.~\ref{fig: photon v2 thermal}(a) we see that the larger switching
temperature $T_s$ in eosL only moves the elliptic-flow peak towards
higher $p_T$. However, as shown in Fig.~\ref{fig: photon v2 thermal}(b), with the TRG rates 
there is factor of two difference in the maximum value between the eosL170 and eosL200 cases. 
This systematics can be deduced from Fig.~\ref{fig: rate comparison} and hydrodynamical evolution as follows:
When the system reaches the cross-over region near the QCD phase transition, there has already been a significant anisotropy developed in the transverse flow which is directly reflected to the thermal photon $v_2$. Thus, if the photon emission is increased (decreased) in this region, the thermal photon $v_2$ increases (decreases). As seen in Fig.~\ref{fig: rate comparison}, with the TRG rates the QGP emission rate is larger than the HG rate. Hence increasing the switching temperature decreases the emission and thus the $v_2$. This effect is seen in Fig.~\ref{fig: photon v2 thermal}(b). With the R92 rates in Fig.~\ref{fig: photon v2 thermal}(a) the situation is slightly different, as in the cross-over region $T=170...200$~MeV the QGP emission rate is larger than the HG rate at small energies and vice versa at high energies. Thus, at low-$p_T$ in 
Fig.~\ref{fig: photon v2 thermal}(a) the situation is similar to panel b (i.e. $v_2$ is larger for eosL170 than for eosL200). At $p_T>2$~GeV, the increase of the switching temperature now increases the total emission and thus making $v_2$ larger for eosL200 than for eosL170.

\begin{figure}[h]
  \includegraphics[height=9.0cm]{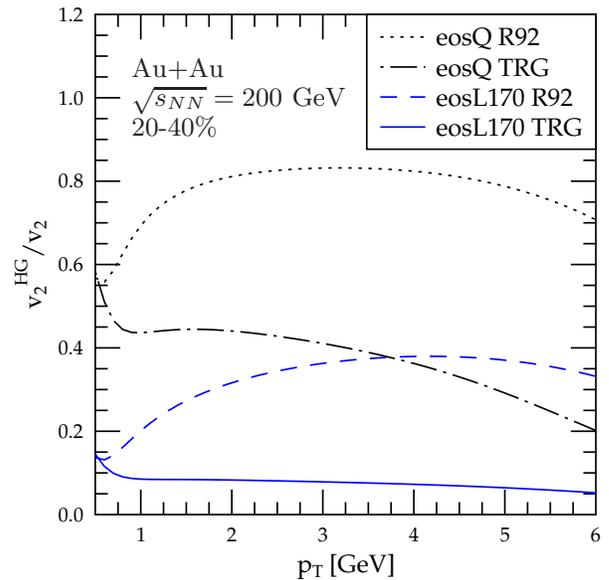}
  \caption{\protect\small (Color online) The contribution to the thermal photon elliptic flow from
  the hadron gas in Au+Au collisions at $\sqrt{s_{NN}} = 200 \GeV$.}
  \label{fig: photon v2 hrg}
\end{figure}

We also notice from the eosQ results in Fig.~\ref{fig: photon v2 thermal} that the maximum $v_2$ decreases by a factor 2 when replacing the R92 rates by the TRG rates. Since the QGP rates in both cases are the same, this signals to us that the hadron gas indeed plays an important role in generating the thermal photon $v_2$ in the eosQ case. To quantify this statement, we have plotted in Fig.~\ref{fig: photon v2 hrg} the fraction of photon $v_2(p_T)$ coming from the HG phase. We define 
$v_2^{\text{HG}}$ as
\begin{equation}
  v_2^{\text{HG}} = \frac{ \int d\phi \cos(2\phi) \frac{dN^{\text{HG}}}{dp_T^2 d\phi dy} }{ \int d\phi \frac{dN^{\text{QGP+HG}}}{dp_T^2 d\phi dy} },
\end{equation}
i.e. relative to to all thermal photons.
We see that the photon $v_2$ can be mostly from the HG (eosQ with R92) or mostly from
the QGP (eosL170 with TRG), or between these extremes (eosQ+TRG and eosL170+R92). Thus, both the EoS and the HG emission rate have a big effect on where the thermal photon $v_2$ originates from.

\begin{figure}[tbh]
  \includegraphics[height=9.0cm]{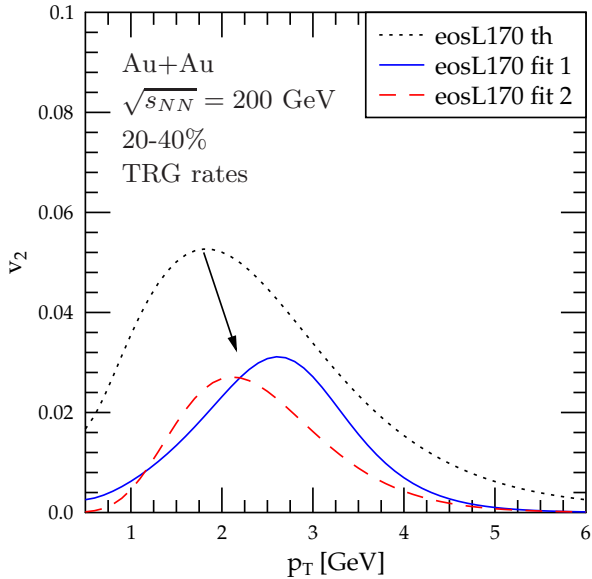}
  \caption{\protect\small (Color online) Elliptic flow of thermal and
  direct photons in Au+Au collisions at $\sqrt{s_{NN}} = 200 \GeV$.}
  \label{fig: photon v2 fits}
\end{figure}

Figure~\ref{fig: photon v2 fits} illustrates how much elliptic flow of thermal photons is washed away if we include other direct photon components assuming that they are produced isotropically. We can see that the final photon $v_2$, obtained from Eq.~\eqref{eq:total_v2} based on the fits 1 and 2,
is clearly smaller than the thermal one, and also that the different fit functions modify the place and shape of the peak, keeping however the maximum $v_2$ roughly the same. 

\begin{figure}[h]
  \includegraphics[width=9.0cm]{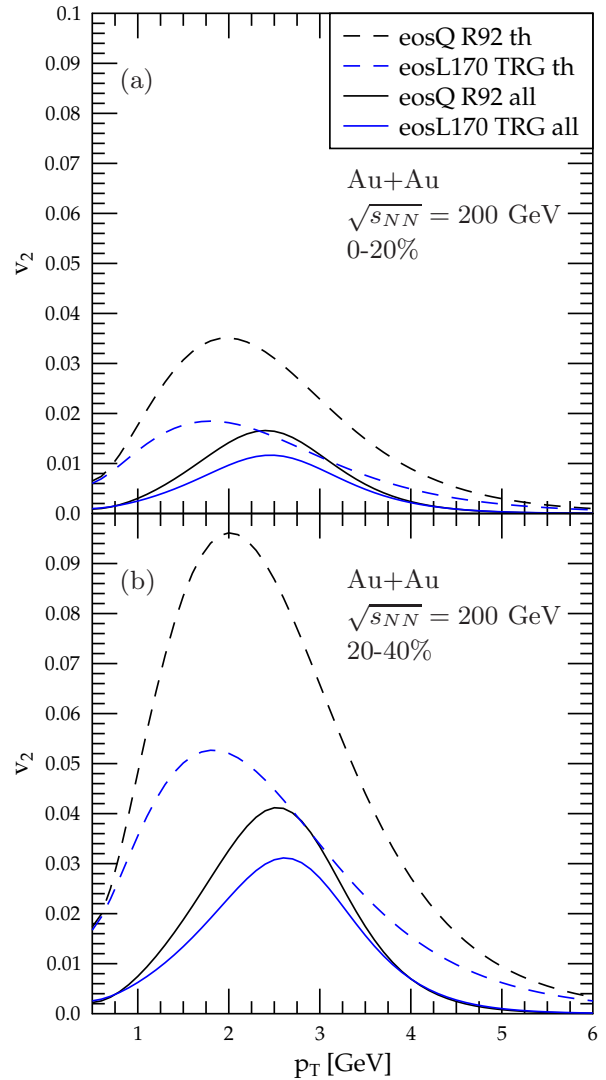}
  \caption{\protect\small (Color online) Elliptic flow of
  direct photons in Au+Au collisions at $\sqrt{s_{NN}} = 200 \GeV$.}
  \label{fig: photon v2 final}
\end{figure}

Finally, in Fig.~\ref{fig: photon v2 final}, we have plotted both the thermal photon and the full direct photon
elliptic flow in 0-20\% and 20-40\% centrality classes for the eosQ+R92 and eosL170+TRG cases. 
The latter can be considered as a state of the art calculation in that a realistic EoS and latest rates are utilized.
In the eosQ+R92 case, the elliptic flow is as large as it can be in our approach. For both cases, the fit 1 is used
to estimate how the other components reduce the elliptic flow. As seen in Fig.~\ref{fig: photon spectra RHIC}, the thermal photon yield is smaller in the eosQ case, and hence the other direct photon components wash away more of the elliptic flow in the eosQ case than in eosL170 case.

\section{Prediction for the LHC}

\begin{figure}[h]
  \includegraphics[height=9.0cm]{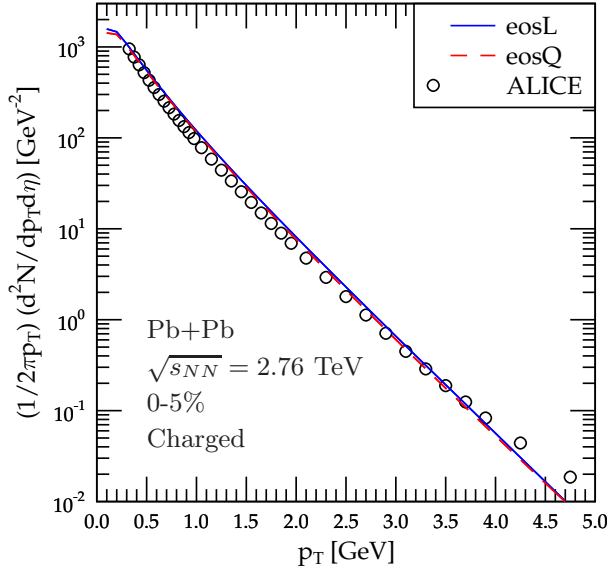}
  \caption{\protect\small (Color online) Transverse momentum spectra of
  charged hadrons in 0-5 \% most central Pb+Pb collisions at $\sqrt{s_{NN}} = 2.76$~TeV.
  Data from ALICE Collaboration \cite{Aamodt:2010jd}.}
  \label{fig: spectra lhc}
\end{figure}

Next, we extrapolate our hydrodynamical modeling to the $\sqrt{s_{NN}} = 2.76$~TeV Pb+Pb collisions at the LHC.
We choose the same sWN initial density profile and decoupling temperatures as at RHIC, and, as explained in Sec.~II.B, use the measured charged-hadron multiplicity \cite{Aamodt:2010pb} to fix the initial entropy and initial time through the EKRT-model (for details, see Ref.~\cite{Renk:2011gj}). As seen in Fig.~\ref{fig: spectra lhc}, a reasonable agreement with the measured charged-hadron $p_T$ spectrum  \cite{Aamodt:2010jd} follows up to $p_T \sim 4$~GeV.
This ensures that our thermal photon calculations are meaningful also at the LHC energies.

\begin{figure}[h!]
  \includegraphics[height=9.0cm]{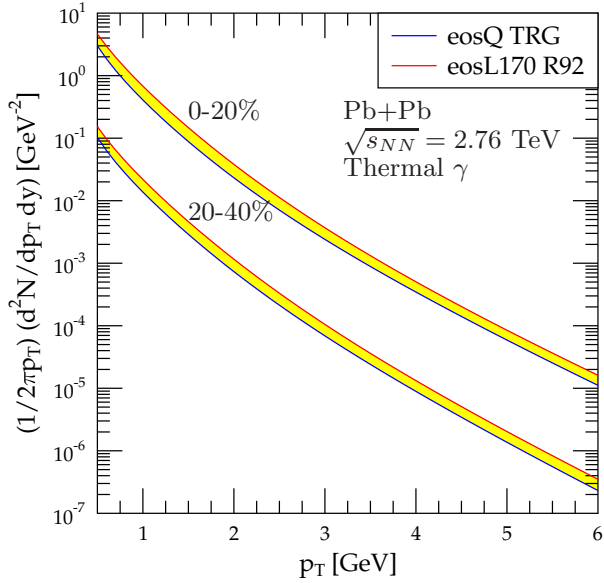}
  \caption{\protect\small (Color online) Transverse momentum spectra of
  thermal photons in Pb+Pb collisions at $\sqrt{s_{NN}} = 2.76$~TeV.}
  \label{fig: photon spectra lhc}
\end{figure}

In Fig.~\ref{fig: photon spectra lhc} we have plotted our prediction for the thermal photon $p_T$ spectrum in the $\sqrt{s_{NN}} = 2.76$~TeV Pb+Pb collisions at the LHC. The bands shown are defined
by the cases  eosL170+R92 and eosQ+TRG, which give the largest and smallest yields, correspondingly.
We see that the uncertainty coming from the EoS and from the HG emission rates is at
largest of the order of 40\%. We note that these predictions are qualitatively quite similar to the predictions given in \cite{Arleo:2004gn,Dusling:2009ej} for $\sqrt{s_{NN}} = 5.5$~TeV.

\begin{figure}[h!]
  \includegraphics[height=9.0cm]{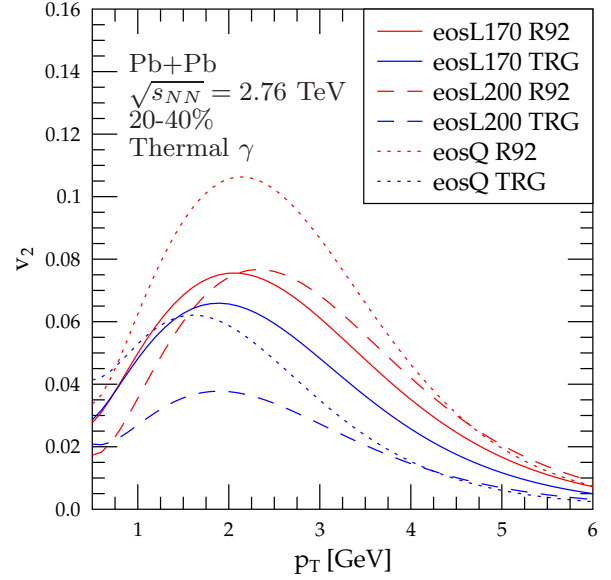}
  \caption{\protect\small (Color online) Elliptic flow of thermal
  photons in Pb+Pb collisions at $\sqrt{s_{NN}} = 2.76$~TeV.}
  \label{fig: photon v2 lhc}
\end{figure}

\begin{figure}[h!]
  \includegraphics[height=9.0cm]{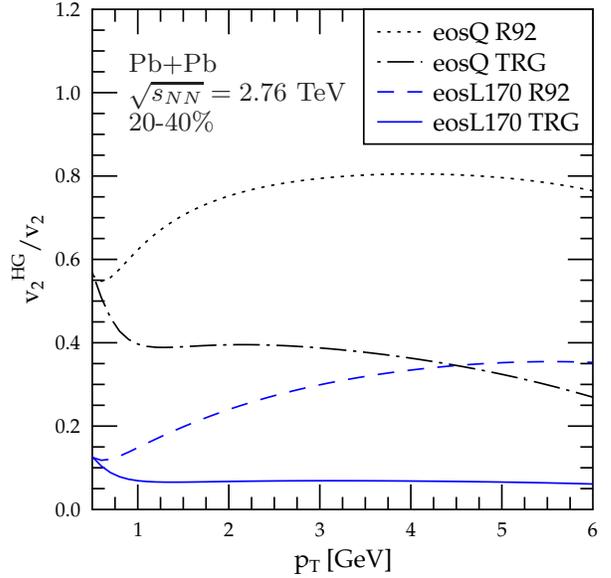}
  \caption{\protect\small (Color online) The contribution to the thermal photon elliptic flow from
  the hadron gas in Pb+Pb collisions at $\sqrt{s_{NN}} = 2.76$~TeV.}
  \label{fig: photon v2 hrg_LHC}
\end{figure}

Since currently we do not have the measured total direct photon spectrum at the LHC available yet (like we had at RHIC),
we can consider only the thermal photon elliptic flow here. It is, however, very interesting to compare the thermal photon elliptic flow with the RHIC results. We have plotted the obtained thermal photon $v_2$ in
Fig.~\ref{fig: photon v2 lhc} for 20-40\% central collisions. We can see that, similarly to the hadronic case \cite{Aamodt:2010pa}, the thermal photon elliptic flow is very similar at RHIC and LHC. This is a non-trivial result since the temperature-range, flow-range as well as the volume factors for photon emission (from the QGP in particular) are larger at the LHC than at RHIC, and, as discussed in \cite{Niemi:2008ta} also the flow asymmetry near the phase transition region is larger at the LHC. Then, when going from RHIC to LHC, in order to arrive at a similar $v_2$ in both cases, the increased flow asymmetry in the numerator of Eq.~\ref{eq: v_2} is compensated by the increased photon yields in the denominator of Eq.~\ref{eq: v_2}. 
In Fig.~\ref{fig: photon v2 hrg_LHC} we again plot the hadronic fraction of $v_2(p_T)$ for the same cases as in Fig.~\ref{fig: photon v2 hrg}. The figure shows that $v_2^{\rm HG}$ is again very close to the corresponding fraction at RHIC.

\section{Discussion}

We have considered the sensitivity of thermal photon production to the EoS and emission rates in heavy-ion collisions at RHIC and LHC.  We have compared the obtained thermal photon yields with the PHENIX measurements, and shown that in the window $2\lesssim p_T\lesssim 3$~GeV the thermal contribution, computed with a realistic EoS (eosL) and latest emission rates (TRG), is reasonably close to the data.  Like in most previous hydrodynamical studies, in the region $p_T \sim 1$~GeV, however, we get a clearly smaller yield than what is measured most recently. We have shown that around $p_T\sim 2$~GeV the thermal photon elliptic flow peaks at a fairly large value, 5 \% in the 20-40\% centrality class with eosL+TRG, but also that the possible other components may wash even half of this away. We emphasize, however, that the amount of $v_2$ wash-out depends on the thermal photon contribution relative to the other components.  Thermal photon production near $p_T\sim 2$~GeV at RHIC  can be expected to increase further once the event-by-event QCD-matter density fluctuations are accounted for \cite{Chatterjee:2011dw}, in which case the thermal photon production can become dominant and the $v_2$ wash-out to decrease or even vanish. 

Constraining our hydrodynamical modeling with the measured charged-hadron spectrum in  $\sqrt{s_{NN}} = 2.76$~TeV Pb+Pb collisions at the LHC, we have predicted the thermal photon $p_T$ spectra and $v_2$. According to our results, elliptic flow of thermal photons at the LHC and RHIC are very similar in the few-$p_T$ region.  For the determination of a possible thermal photon window, and consequently thermal photon $v_2$, it will be extremely interesting to see the direct photon data at the LHC.

Next, one should consider the effects of event-by-event density fluctuations \cite{Andrade:2008xh,Petersen:2010md,Werner:2010aa,Holopainen:2010gz,Schenke:2010rr,Qiu:2011iv} on thermal photon elliptic flow. 
On the theoretical side, one would need a better understanding of how the degrees of freedom in the QGP would be best accounted for when computing thermal photon production, as well as a better control over the photon emission in the phase transition region. Also the dissipative hydrodynamical effects to thermal photon production should be studied further, so far only the very first steps into this direction have been taken, see Ref.~\cite{Dusling:2009bc}.

\begin{acknowledgments} 
We gratefully acknowledge financial support by the Academy of Finland, KJE's project 133005. In addition, HH is supported by the national Graduate School of Particle and Nuclear Physics. We acknowledge CSC -- IT Center for Science in Espoo, Finland, for the allocation of computational resources. We would like to thank R. Chatterjee for useful discussions. 
\end{acknowledgments}


\begin{thebibliography}{99}

\bibitem{Adler:2006hu}
  S.~S.~Adler {\it et al.} [ PHENIX Collaboration ],
  Phys.\ Rev.\ Lett.\  {\bf 96 } (2006)  202301.

\bibitem{Aurenche:2006vj}
  P.~Aurenche, M.~Fontannaz, J.~-P.~Guillet, E.~Pilon, M.~Werlen,
  Phys.\ Rev.\  {\bf D73 } (2006)  094007.

\bibitem{Arleo:2011gc}
  F.~Arleo, K.~J.~Eskola, H.~Paukkunen, C.~A.~Salgado,
  JHEP {\bf 1104 } (2011)  055.

\bibitem{Eskola:2009uj}
  K.~J.~Eskola, H.~Paukkunen, C.~A.~Salgado,
  JHEP {\bf 0904 } (2009)  065.

\bibitem{Fries:2002kt}
  R.~J.~Fries, B.~Muller, D.~K.~Srivastava,
  Phys.\ Rev.\ Lett.\  {\bf 90 } (2003)  132301.

\bibitem{Fries:2005zh}
  R.~J.~Fries, B.~Muller, D.~K.~Srivastava,
  Phys.\ Rev.\  {\bf C72 } (2005)  041902.

\bibitem{Turbide:2007mi}
  S.~Turbide, C.~Gale, E.~Frodermann, U.~Heinz,
  Phys.\ Rev.\  {\bf C77 } (2008)  024909.

\bibitem{Liu:2008eh}
  F.~-M.~Liu, T.~Hirano, K.~Werner, Y.~Zhu,
  Phys.\ Rev.\  {\bf C79 } (2009)  014905.

\bibitem{Qin:2009bk}
  G.~-Y.~Qin, J.~Ruppert, C.~Gale, S.~Jeon, G.~D.~Moore,
  Phys.\ Rev.\  {\bf C80 } (2009)  054909.

\bibitem{Huovinen:1998tq}
  P.~Huovinen, P.~V.~Ruuskanen, J.~Sollfrank,
  Nucl.\ Phys.\  {\bf A650 } (1999)  227-244.

\bibitem{Huovinen:2001wx}
  P.~Huovinen, P.~V.~Ruuskanen, S.~S.~R\"as\"anen,
  Phys.\ Lett.\  {\bf B535 } (2002)  109-116.

\bibitem{Rasanen:2002qe}
  S.~S.~R\"as\"anen,
  Nucl.\ Phys.\  {\bf A715 } (2003)  717-725.

\bibitem{d'Enterria:2005vz}
  D.~G.~d'Enterria, D.~Peressounko,
  Eur.\ Phys.\ J.\  {\bf C46 } (2006)  451-464.

\bibitem{Chatterjee:2005de}
R.~Chatterjee, E.~S.~Frodermann, U.~W.~Heinz, D.~K.~Srivastava,
  Phys.\ Rev.\ Lett.\  {\bf 96 } (2006)  202302.

\bibitem{Chatterjee:2008tp}
  R.~Chatterjee, D.~K.~Srivastava,
  Phys.\ Rev.\  {\bf C79 } (2009)  021901.

\bibitem{Liu:2009kta}
  F.~-M.~Liu, T.~Hirano, K.~Werner, Y.~Zhu,
  Phys.\ Rev.\  {\bf C80 } (2009)  034905.

\bibitem{Adler:2006yt}
  S.~S.~Adler {\it et al.} [ PHENIX Collaboration ],
  Phys.\ Rev.\ Lett.\  {\bf 98 } (2007)  012002.

\bibitem{Adler:2005ig}
  S.~S.~Adler {\it et al.} [ PHENIX Collaboration ],
  Phys.\ Rev.\ Lett.\  {\bf 94 } (2005)  232301.

\bibitem{:2008fqa}
  A.~Adare {\it et al.} [ PHENIX Collaboration ],
  Phys.\ Rev.\ Lett.\  {\bf 104 } (2010)  132301.

\bibitem{Arleo:2004gn}
F.~Arleo {\em et~al.}, {\em Photon Physics in Heavy Ion Collisions}, CERN
Yellow Book report (2004), hep-ph/0311131.

\bibitem{Dusling:2009ej}
  K.~Dusling, I.~Zahed,
  Phys.\ Rev.\  {\bf C82 } (2010)  054909.

\bibitem{Eskola:1999fc}
  K.~J.~Eskola, K.~Kajantie, P.~V.~Ruuskanen, K.~Tuominen,
  Nucl.\ Phys.\  {\bf B570 } (2000)  379-389.

\bibitem{Eskola:2005ue}
  K.~J.~Eskola, H.~Honkanen, H.~Niemi, P.~V.~Ruuskanen, S.~S.~Rasanen,
  Phys.\ Rev.\  {\bf C72 } (2005)  044904.

\bibitem{Niemi:2008ta}
  H.~Niemi, K.~J.~Eskola, P.~V.~Ruuskanen,
  Phys.\ Rev.\  {\bf C79 } (2009) 024903.

\bibitem{Aamodt:2010pb}
  K.~Aamodt {\it et al.} [ The ALICE Collaboration ],
  Phys.\ Rev.\ Lett.\  {\bf 105 } (2010)  252301.

\bibitem{Renk:2011gj}
  T.~Renk, H.~Holopainen, R.~Paatelainen, K.~J.~Eskola,
  arXiv:1103.5308 [hep-ph].

\bibitem{:2008ez}
  B.~I.~Abelev {\it et al.} [ STAR Collaboration ],
  Phys.\ Rev.\  {\bf C79 } (2009)  034909.

\bibitem{Kolb:2001qz}
  P.~F.~Kolb, U.~W.~Heinz, P.~Huovinen, K.~J.~Eskola, K.~Tuominen,
  Nucl.\ Phys.\  {\bf A696 } (2001)  197-215.

\bibitem{Boris}
  J.~P. Boris, D.~L. Book,
  J. Comput. Phys. {\bf A11} (1973) 38.

\bibitem{Zalesak}
  S.~T. Zalesak,
  J. Comput. Phys. {\bf A31} (1979) 248.

\bibitem{Sollfrank:1996hd}
  J.~Sollfrank, P.~Huovinen, M.~Kataja, P.~V.~Ruuskanen, M.~Prakash, R.~Venugopalan,
  Phys.\ Rev.\  C {\bf 55} (1997) 392.

\bibitem{Laine:2006cp}
  M.~Laine, Y.~Schroder,
  Phys.\ Rev.\  {\bf D73 } (2006)  085009.

\bibitem{Huovinen:2009yb}
  P.~Huovinen, P.~Petreczky,
  Nucl.\ Phys.\  {\bf A837 } (2010)  26-53.

\bibitem{Cooper}
  F.~Cooper, G.~Frye,
  Phys. Rev. {\bf D10} (1974) 186.

\bibitem{Arnold:2001ba}
  P.~B.~Arnold, G.~D.~Moore, L.~G.~Yaffe,
  JHEP {\bf 0111 } (2001)  057.

\bibitem{Arnold:2001ms}
  P.~B.~Arnold, G.~D.~Moore, L.~G.~Yaffe,
  JHEP {\bf 0112 } (2001)  009.

\bibitem{Karsch:1987kz}
  F.~Karsch,
  Z. Phys. {\bf C38} (1988) 147.

\bibitem{Kapusta:1991qp}
  J.~I. Kapusta, P.~Lichard, D.~Seibert,
  Phys. Rev. {\bf D44} (1991) 2774.

\bibitem{Xiong:1992ui}
  L.~Xiong, E.~V.~Shuryak, G.~E.~Brown,
  Phys.\ Rev.\  {\bf D46 } (1992)  3798-3801.

\bibitem{Nadeau:1992cn}
  H.~Nadeau, J.~I. Kapusta, P.~Lichard,
  Phys. Rev. {\bf C45} (1992) 3034.

\bibitem{Turbide:2003si}
  S.~Turbide, R.~Rapp, C.~Gale,
  Phys.\ Rev.\  {\bf C69 } (2004)  014903.

\bibitem{Adler:2003cb}
  S.~S.~Adler {\it et al.} [ PHENIX Collaboration ],
  Phys.\ Rev.\  {\bf C69 } (2004)  034909.

\bibitem{:2008ed}
  B.~I.~Abelev {\it et al.} [ STAR Collaboration ],
  Phys.\ Rev.\  {\bf C77 } (2008)  054901.

\bibitem{Holopainen:2010gz}
  H.~Holopainen, H.~Niemi, K.~J.~Eskola,
  Phys.\ Rev.\  {\bf C83 } (2011)  034901.

\bibitem{Romatschke:2007mq}
  P.~Romatschke, U.~Romatschke,
  Phys.\ Rev.\ Lett.\  {\bf 99 } (2007)  172301.

\bibitem{Adare:2009qk}
  A.~Adare {\it et al.} [ PHENIX Collaboration ],
  Phys.\ Rev.\  {\bf C81 } (2010)  034911.

\bibitem{Chatterjee:2011dw}
  R.~Chatterjee, H.~Holopainen, T.~Renk, K.~J.~Eskola,
  arXiv:1102.4706 [hep-ph].

\bibitem{Aamodt:2010jd}
  K.~Aamodt {\it et al.} [ ALICE Collaboration ],
  Phys.\ Lett.\  {\bf B696 } (2011)  30-39.

\bibitem{Aamodt:2010pa}
  K.~Aamodt {\it et al.} [ The ALICE Collaboration ],
  arXiv:1011.3914 [nucl-ex].

\bibitem{Andrade:2008xh}
  R.~P.~G.~Andrade, F.~Grassi, Y.~Hama, T.~Kodama, W.~L.~Qian,
  Phys.\ Rev.\ Lett.\  {\bf 101 } (2008)  112301.

\bibitem{Petersen:2010md}
  H.~Petersen, M.~Bleicher,
  Phys.\ Rev.\  {\bf C81 } (2010)  044906.

\bibitem{Werner:2010aa}
  K.~Werner, I.~Karpenko, T.~Pierog, M.~Bleicher, K.~Mikhailov,
  Phys.\ Rev.\  {\bf C82 } (2010)  044904.

\bibitem{Schenke:2010rr}
  B.~Schenke, S.~Jeon, C.~Gale,
  Phys.\ Rev.\ Lett.\  {\bf 106 } (2011)  042301.

\bibitem{Qiu:2011iv}
  Z.~Qiu, U.~W.~Heinz,
  arXiv:1104.0650 [nucl-th].

\bibitem{Dusling:2009bc}
  K.~Dusling,
  Nucl.\ Phys.\  {\bf A839 } (2010)  70-77.


\end{thebibliography}
\end{document}